\documentclass[aps,superscriptaddress,prl,twocolumn,amsmath]{revtex4-1}
\usepackage{amssymb}
\usepackage{amsmath}
\usepackage{graphicx,subfigure,float}
\usepackage[percent]{overpic}
\usepackage{dcolumn}
\usepackage{bm}
\usepackage{color}
\usepackage{array}
\usepackage{hyperref}
\usepackage[mathlines]{lineno}
\begin{document}
\title{Weyl-type nodal chains in X$_2$MnO$_4$ (X= Li, Na)}
\author{R. R. Kang}
\affiliation{Hunan Provincial Key laboratory of Thin Film Materials and Devices, School of Material Sciences and Engineering, Xiangtan University, Xiangtan 411105, China}
\author{S. D. He}
\affiliation{Hunan Provincial Key laboratory of Thin Film Materials and Devices, School of Material Sciences and Engineering, Xiangtan University, Xiangtan 411105, China}
\author{P. Zhou}
\email{zhoupan71234@xtu.edu.cn}
\affiliation{ School of Material Sciences and Engineering, Xiangtan University, Xiangtan 411105, China}
\author{L. Z. Sun}
\email{lzsun@xtu.edu.cn}
\affiliation{Hunan Provincial Key laboratory of Thin Film Materials and Devices, School of Material Sciences and Engineering, Xiangtan University, Xiangtan 411105, China}
\date{\today}
\begin{abstract}
\indent Recently, magnetic topological semimetals have received a lot of attention due to their potential applications in the field of spintronics. By using first-principles calculations, we propose that two ferromagnetic spinel materials of X$_2$MnO$_4$ (X= Li, Na) have Weyl-type nodal chains around the Fermi level. Their stabilities are validated by cohesive energies, phonon dispersions, and elastic constants. The nodal chains are composed of two types of nodal loops, which are protected by the glide operation $\tilde{\mathcal{M}}_{z}$, the mirror operation ${\mathcal{M}}_{\overline101}$ and their equivalent. The drumhead surface states are observed on the (001) surface and they exhibit nontrivial topological features. In addition, under different electron correlations and lattice strains, the semimetal states of these two materials are well kept. Our work provides two promising candidates for exploring the combination of magnetic materials and topological semimetal states.\\
\end{abstract}
\maketitle
\section*{\uppercase\expandafter{\romannumeral1}. INTRODUCTION}
\indent Topological semimetals \cite{1} in solid systems are newly discovered quantum matter states in recent years, and they are characterized by the existence of stable crossing points around the Fermi level. Based on the degeneracy and distribution of these crossing points in reciprocal space, topological semimetals can be divided into Dirac semimetals \cite{2,3,4,5}, Weyl semimetals \cite{6,7,8,9}, nodal line semimetals \cite{8,10,11,12}, etc. The characteristic of the nodal line semimetal is that the energy band crossing points form closed or opened lines. When multiple nodal lines coexist in the Brillouin zone (BZ), they can form various configurations, such as nodal net \cite{13,14,15}, nodal box \cite{16,17}, Hopf link \cite{19,20,21}, or nodal chain \cite{50,51,52,53}. A nodal chain semimetal contains a chain of connected loops in momentum space, and they are protected by certain crystal symmetries.\\
\indent Recently, magnetic topological semimetals are getting more and more attention \cite{54,55,56,57,58}. First, there are abundant unconventional interactions between magnetism and topology. Moreover, magnetic topological semimetal is a promising platform for studying the interaction between topological semimetal states and electronic correlations. Besides, to date, the quantum anomalous Hall effect has been observed in some systems, such as magnetically-doped topological insulator thin films \cite{60,61,62} and van der Waals layered magnetic topological insulator MnBi$_2$Te$_4$ \cite{63,64,65}. However, the experiments can only be carried out at extremely low temperatures, which severely limits its future practical applications \cite{22}. Magnetic topological semimetals can provide a new way to realize the quantum anomalous Hall effect at higher temperatures. Nevertheless, ferromagnetic semimetals are uncommon in nature, and magnetic topological semimetals are rarer. Researchers have theoretically proposed some magnetic topological semimetals, like HgCr$_2$Se$_4$ \cite{25}, Co$_2$TiX (X=Si, Ge, Sn) \cite{23}, Fe$_3$O$_4$ \cite{24}. However, only a few of them have been confirmed in experiments, such as Co$_2$MnCa \cite{26,27}. So it is very important to explore more magnetic topological semimetal materials for studying their fascinating properties.\\
\indent In this paper, we predict two magnetic topological semimetal materials X$_2$MnO$_4$ (X= Li, Na). Through analyzing the distribution of the crossing points, we confirm these two materials are Weyl-type nodal chain semimetals and the chains are formed by two kinds of inequivalent nodal loops. Moreover, the node chains are robust to electron correlations and lattice strains. After considering SOC, only a small band gap is opened for all crossing points of the nodal chains. Our findings will facilitate the experimental studies on spin-polarized nodal chains towards both fundamental discoveries and potential spintronics applications.\\
\section*{\uppercase\expandafter{\romannumeral2}. COMPUTATIONAL METHODOLOGY}
\indent The first-principles calculations were based on density functional theory (DFT) \cite{28} as implemented in the Vienna ab initio simulation package (VASP) \cite{29}. The generalized gradient approximation (GGA) with Perdew-Burke-Ernzerhof (PBE) \cite{30} was used to simulate the exchange-correlation potential. The all-electron projector augmented wave (PAW) method \cite{31} was used to describe the interaction between atomic nuclei and electrons. In our calculations, the cutoff energies were set at 520 eV, and the energies and forces were converged to $10^{-6}$ eV and 0.05 eV/{\AA}, respectively. The BZ was sampled with a 11 $\times$ 11 $\times$ 11 Monkhorst-Pack k-point grid. In addition, considering the possible strong correlation effects for the d orbitals of Mn atoms, the DFT + U \cite{32} was used in our calculations (U= 4 eV). The phonon spectrum were obtained from density functional perturbation theory (DFPT) \cite{33}. The surface states were calculated from WannierTools \cite{34}.\\
\begin{figure}
	\centering
	\includegraphics[trim={0.0in 0.0in 0.0in 0.0in},clip,width=3.5in]{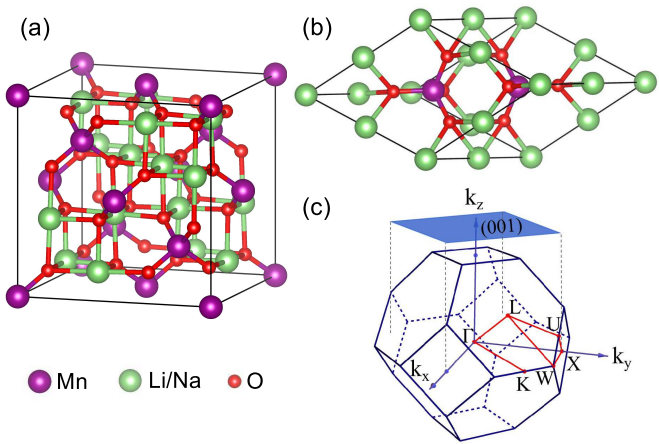}
	\caption{(a) Conventional and (b) primitive cells of X$_2$MnO$_4$ (X= Li, Na). The purple, green, and red spheres represent Mn, Li/Na, and O atoms, respectively. (c) Three-dimensional BZ in which the high-symmetry paths are highlighted in red lines and the area shaded in blue corresponds to the projected (001) surface.}
	\label{structure}
\end{figure}
\section*{\uppercase\expandafter{\romannumeral3}. CRYSTAL STRUCTURE}
\indent As shown in Figs. 1(a) and 1(b), X$_2$MnO$_4$ (X= Li, Na) crystals we proposed belong to the spinel structure with face-centered cubic lattice. Their space group is Fd-3m (No. 227) including the glide mirror $\tilde{\mathcal{M}}_{z}$ : (x, y, z) $\rightarrow$ (x + 1/4, y + 3/4, -z + 1/2) and mirror ${\mathcal{M}}_{\overline101}$ : (x, y, z) $\rightarrow$ (z, y, x). The Mn and alkali metal atoms are located in the centers of the O tetrahedron and octahedron, respectively. The fully optimized lattice constants are 5.81 {\AA} for Li$_2$MnO$_4$ and 6.26 {\AA} for Na$_2$MnO$_4$. To obtain the ground magnetic states for them, we calculated the total energies of different magnetic states: nonmagnetic (NM), ferromagnetic (FM) and antiferromagnetic (AFM), as listed in Table I. It is found that the energy of FM is the lowest, which indicates their magnetic ground state is FM state. We further calculated their Bader charges to obtain the charge transfer. In Li$_2$MnO$_4$ (Na$_2$MnO$_4$), each Li (Na) and Mn atom lose about 0.89 (0.81) $|$e$|$ and 1.79 (2.02) $|$e$|$, respectively, and each O atom gains about 0.89 (0.91) $|$e$|$.\\
\begin{table}
\renewcommand\arraystretch{2}
  \centering
  \caption{Total energy E$_{tot}$ per unit cell (in eV, relative to that of the FM ground state) for X$_2$MnO$_4$ (X= Li, Na) under different magnetic configurations. The values are obtained by the GGA + U method with U = 4.0 eV.}\label{table1}
  \begin{tabular}{p{2.1cm} p{2.0cm}<{\centering} p{2.0cm}<{\centering} p{2.0cm}<{\centering}}
  \hline\hline
    System & NM & FM & AFM \\
  \hline
    Li$_2$MnO$_4$ & 1.6207 & 0.0 & 0.4263 \\
    Na$_2$MnO$_4$ & 1.0795 & 0.0 & 0.3664 \\
  \hline\hline
  \end{tabular}
\end{table}\\
\section*{\uppercase\expandafter{\romannumeral4}. STABILIY}
\indent Because these two materials have not been synthesized, we evaluated their cohesive energies, dynamic and mechanical stabilities. The cohesive energies $E_{coh}$ are defined as
$$E_{coh} = (4E_{Li/Na} + 2E_{Mn} + 8E_{O} - E_{X_2MnO_4})/14,$$
where E$_{X_2MnO_4}$ represents the total energy of the corresponding primitive cell, E$_{Li/Na}$, E$_{Mn}$ and E$_O$ are the energies of single Li/Na, Mn and O atoms, respectively. Their cohesive energies are 4.13 eV/atom and 3.97 eV/atom, respectively. These values are comparable to the corresponding values of experimentally synthesized Na$_2$MoO$_4$ (4.70 eV/atom) and Mg$_2$CoO$_4$ (4.77 eV/atom)\cite{35}. These results suggest that they are energetically stable and are expected to be synthesized experimentally. The phonon dispersions of the two compounds are shown in Figs. 2(a) and 2(b), respectively. No imaginary frequencies are observed, which indicates that they are dynamically stable at their ground state. We estimate the mechanical properties of X$_2$MnO$_4$ based on the energy-strain method. There are three independent elastic constants for a cubic crystal, including C$_{11}$, C$_{12}$ and C$_{44}$. The calculated values for Li$_2$MnO$_4$ (Na$_2$MnO$_4$) are 125.22, 49.04 and 41.12 Gpa (115.65, 31.94 and 25.09 Gpa), respectively. The mechanical stability criteria of a cubic crystal is as follows\cite{36}:
$$\begin{aligned}
&C _{11}-C _{12}>0, \\
&C _{44}>0, \\
&C _{11}+2 C _{12}>0.
\end{aligned}$$\\
\indent Both of the two materials comply with the above criteria, indicating that they are mechanically stable.
\begin{figure*}
	\centering
	\includegraphics[trim={0.0in 0.0in 0.0in 0.0in},clip,width=5.6in]{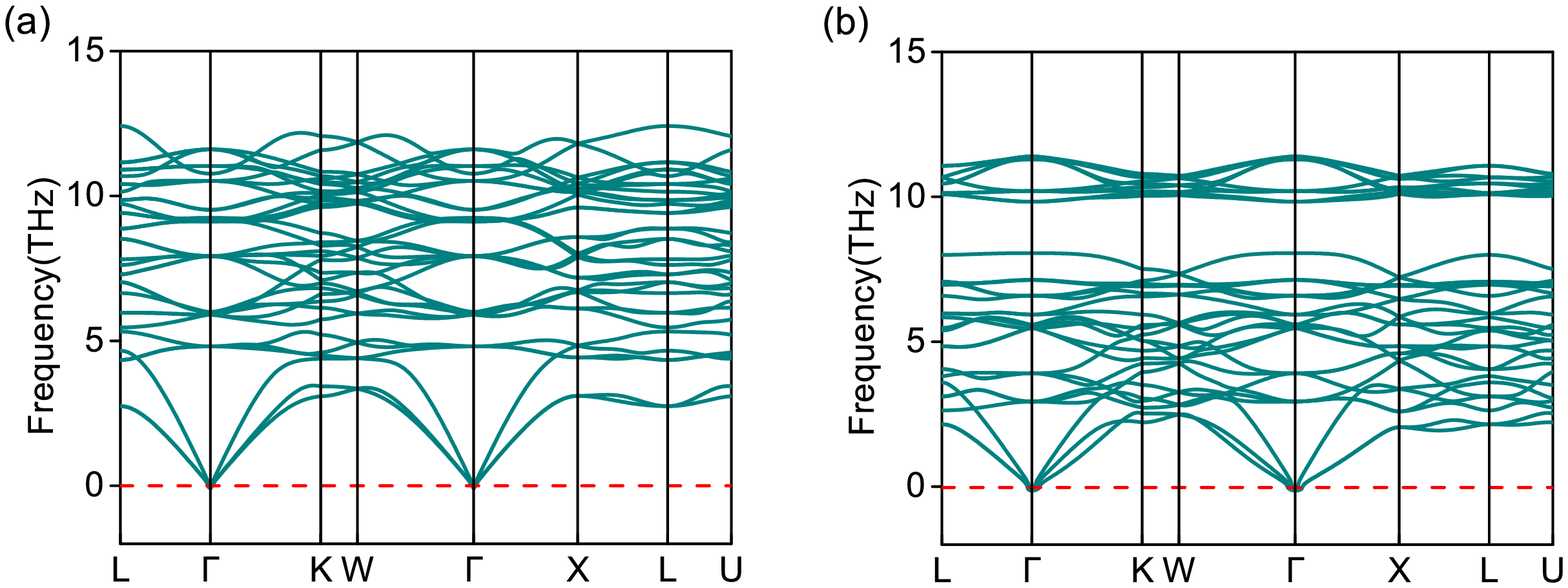}
	\caption{Phonon dispersions of (a) Li$_2$MnO$_4$ and (b) Na$_2$MnO$_4$.}
	\label{phon}
\end{figure*}
\begin{figure*}
	\centering
	\includegraphics[trim={0.0in 0.0in 0.0in 0.0in},clip,width=6.5in]{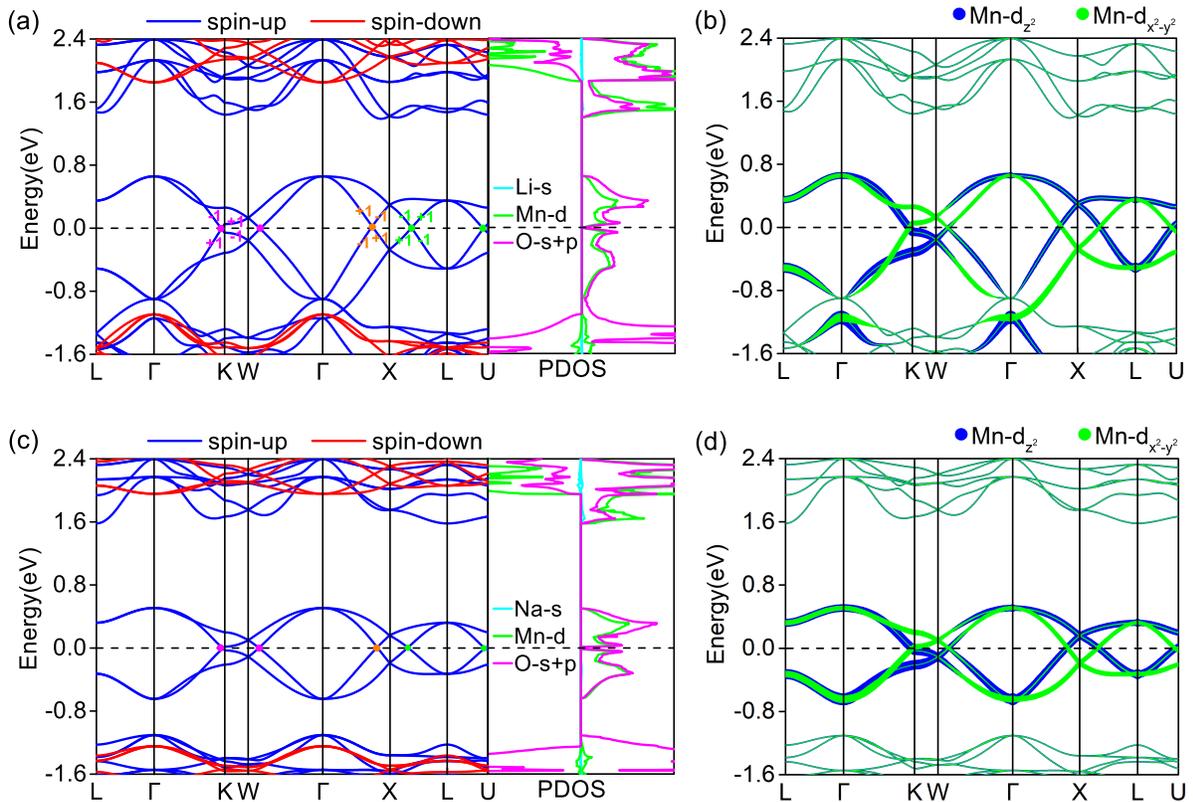}
	\caption{Spin-polarized electronic band structures of (a) Li$_2$MnO$_4$ and (c) Na$_2$MnO$_4$ in the absence of SOC. The corresponding PDOS are also shown. The crossing points denoted by the magenta (green) dots are on the plane of $\Gamma$-K-W-X ($\Gamma$-X-U-L) and are protected by the mirror $\tilde{\mathcal{M}}_{z}$ (${\mathcal{M}}_{\overline101}$), whereas the orange point is shared by two mirrors. The projected band structures of Mn-d$_{z^2}$/d$_{x^2-y^2}$ orbitals for (b) Li$_2$MnO$_4$ and (d) Na$_2$MnO$_4$.}
	\label{banddos}
\end{figure*}
\begin{figure*}
	\centering
	\includegraphics[trim={0.0in 0.0in 0.0in 0.0in},clip,width=5.6in]{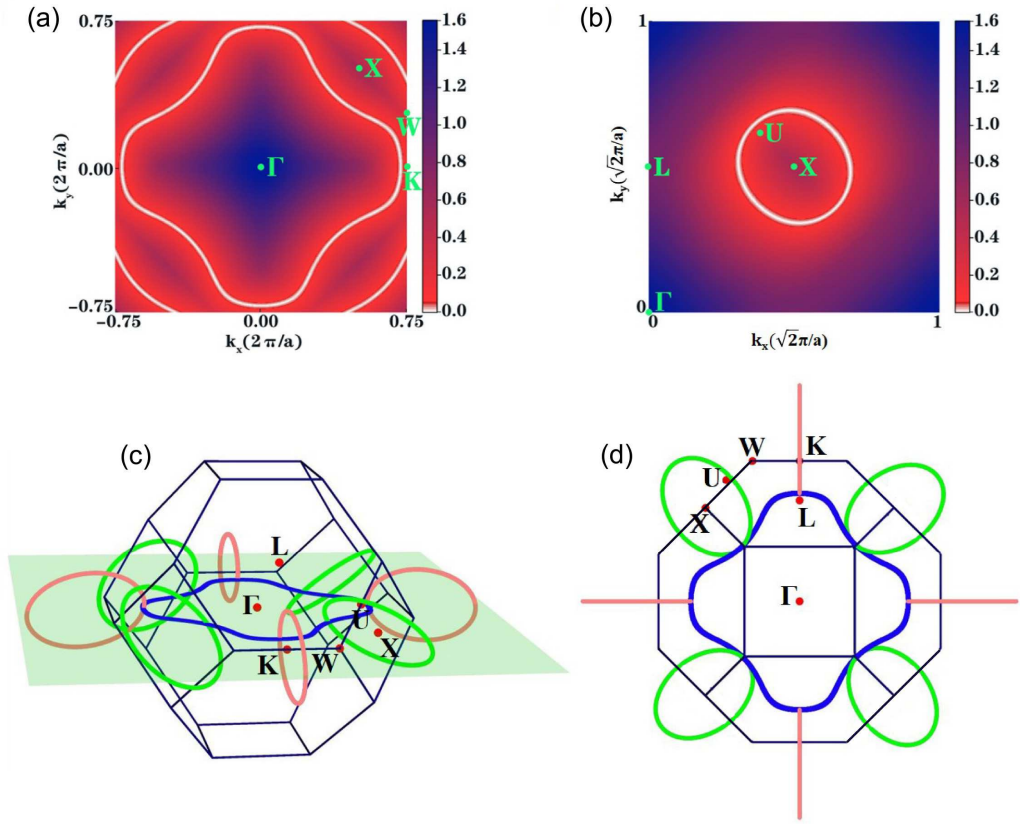}
	\caption{The shapes of (a) the NL1 and (b) the NL2 for Li$_2$MnO$_4$. The color map indicates the local gap between two crossing bands. (c) A schematic diagram of the connection between NL1 (blue loop) and NL2 (green loops), where the pink loops are the equivalences of the green loops. (d) Top views of (c).}
	\label{pband}
\end{figure*}
\begin{figure*}
	\centering
	\includegraphics[trim={0.0in 0.0in 0.0in 0.0in},clip,width=6.8in]{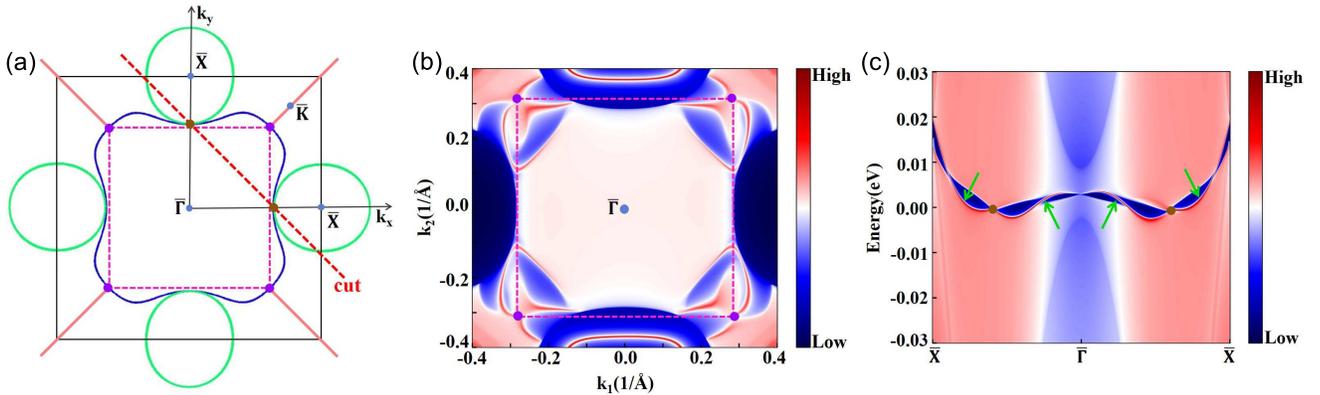}
	\caption{(a) Projection of the nodal networks on the (001) surface, and the connecting points between different nodal loops along the $\Gamma$-K and $\Gamma$-X directions are marked by purple and brown dots, respectively. (b) The (001) surface states of the connecting points in the $\Gamma$-K directions for Li$_2$MnO$_4$ at the Fermi level. (c) Surface band structures along the red dotted line (cut) indicated in (a), where the green arrows point to the drumhead-like surface states.}
	\label{pband}
\end{figure*}
\begin{figure*}
	\centering
	\includegraphics[trim={0.0in 0.0in 0.0in 0.0in},clip,width=5.6in]{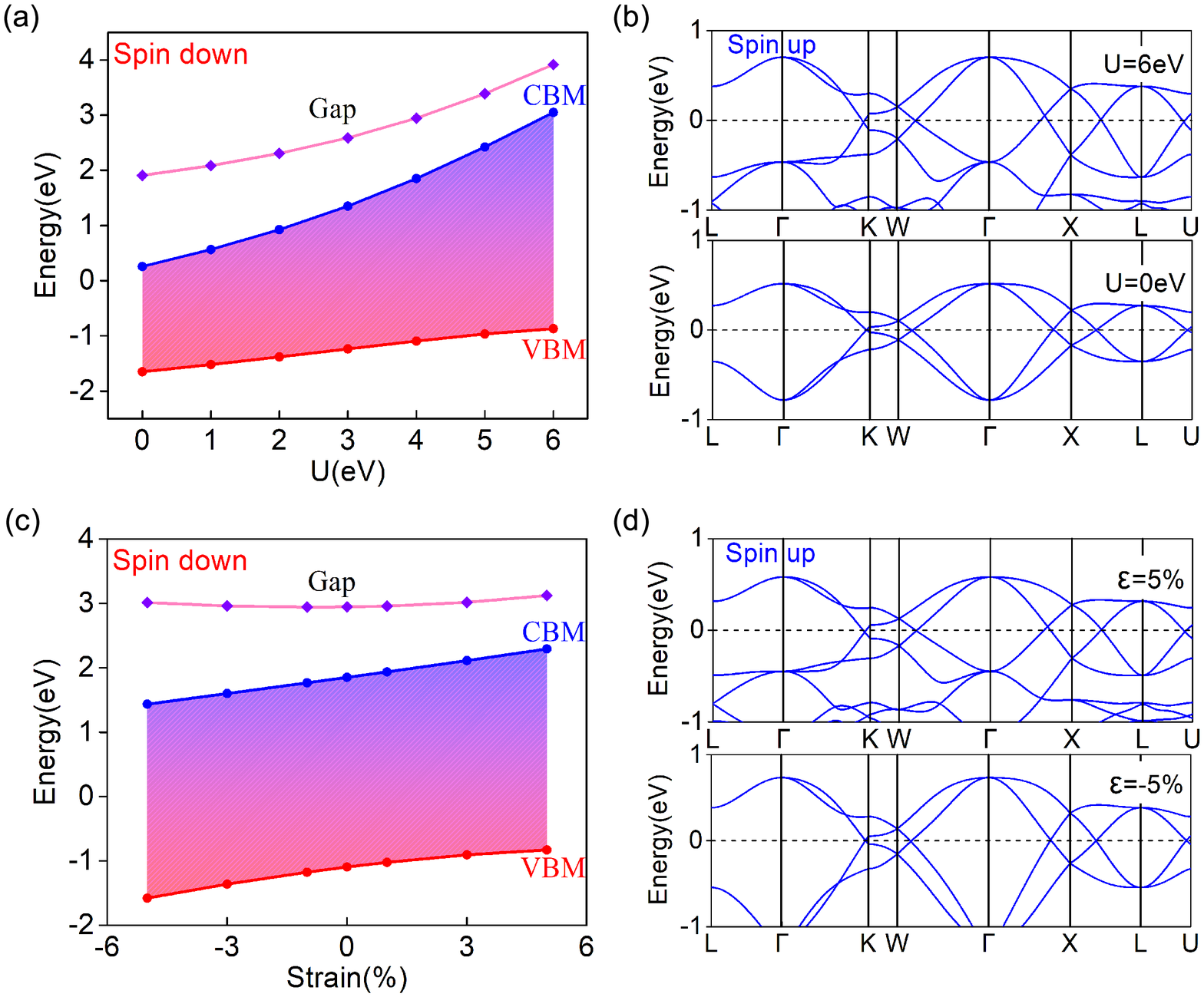}
	\caption{Evolution of spin-down band gaps for Li$_2$MnO$_4$ under different U values (a) and strains (c). Variation curves for valence band maximum (VBM) and conduction band minimum (CBM) are also shown. (b) Spin-up band structures under U = 6 eV and U = 0 eV. (d) Band structures in the spin-up channel with 5\% and -5\% strains.}
\label{Li-UandStrain}
\end{figure*}\\
\section*{\uppercase\expandafter{\romannumeral5}. WEYL-TYPE NODAL CHAIN}
\indent Next, we study the electronic properties of X$_2$MnO$_4$ in the absence of SOC. First of all, we calculated the band structures and projected density of states (PDOS), as shown in Figs. 3(a) and 3(c). The bands with the opposite spin channels are separated from each other. For the spin-up channel, the bands pass through the Fermi level and exhibit metallic features, while the spin-down bands have a big gap (2.94 eV for Li$_2$MnO$_4$ and 3.20 eV for Na$_2$MnO$_4$), which indicates that both of them are half-metal materials. From the PDOS, we can see that the spin-up electronic states around the Fermi level are mainly contributed by the d orbitals of Mn and the s, p orbitals of O atoms. To investigate further which d orbitals contribute to electronic states near the Fermi level, we plot the projected bands of Mn-d orbitals, as shown in Figs. 3(b) and 3(d). We can see that they mainly come from the $d_{z^{2}}$ and $d_{x^{2}-y^{2}}$ orbitals of Mn atoms. This is because, under the crystal field of the O tetrahedron, the d orbitals of Mn atoms split into two groups: the triple degenerate $t_{2}$ ($d_{xy}$, $d_{yz}$ and $d_{xz}$) orbitals with higher energies and the double degenerate $e$ ($d_{z^{2}}$ and $d_{x^{2}-y^{2}}$) orbitals with lower energies. In addition, we find the four bands near the Fermi level are isolated from other bands in energies for Na$_2$MnO$_4$, but not for Li$_2$MnO$_4$.\\
\indent As shown in Figs. 3(a) and 3(c), a series of crossing points can be clearly seen in the paths of $\Gamma$-K, W-$\Gamma$, $\Gamma$-X, X-L and L-U. We analyze the distribution of band crossing points in momentum space. It is fascinating that these crossing points are not isolated, but belong to two different types of nodal loops. More specifically, the band crossings in the $\Gamma$-K, W-$\Gamma$, $\Gamma$-X paths form a node loop in the k$_z$ = 0 plane, denoted as NL1, as shown in Fig. 4(a) for Li$_2$MnO$_4$ [Fig. S1(a) for Na$_2$MnO$_4$]. The $\Gamma$ point is the center of the loop. Because of the cubic symmetry of X$_2$MnO$_4$, there are also nodal loops in the planes of k$_x$ = 0 and k$_y$ = 0, and these nodal loops cross each other to form an inner nodal chain structure\cite{37}. To determine these crossing points are accidental or protected, we calculated the irreducible representations (irreps) of the Bloch states around these crossing points. It is found that the crossing bands belong to the irreps with opposite eigenvalues of $\tilde{\mathcal{M}}_{z}$, as shown in Figs. 3(a) and 3(c), so the NL1 is protected by the glide operation. The band crossings on the $\Gamma$-X, X-L and L-U paths form a nodal loop locating in the $\Gamma$-X-L plane, as shown in Fig. 4(b) for Li$_2$MnO$_4$ [Fig. S1(b) for Na$_2$MnO$_4$]. This nodal loop is denoted as NL2. The shape of it resemble an ellipse centered at point X. The results of irreps reveal NL2 are protected by mirror symmetry ${\mathcal{M}}_{\overline101}$. Besides, according to the cubic symmetry and equivalent mirror operations (${\mathcal{M}}_{\overline101}$ and ${\mathcal{M}}_{1\overline10}$), we find equivalent eight nodal loops can be found around the NL1, as shown by the pink and green loops in Fig. 4(c) and 4(d).\\
\indent To verify the topology nature of the NL1 and NL2, we calculated the Berry phases for a closed loop surrounding each nodal loop as
$$P_{B}=\oint_{L} A(k) \cdot d k$$
where $\mathrm{A}(k)=-i\left\langle\varphi(k)\left|\nabla_{k}\right| \varphi(k)\right\rangle$ is called the Berry vector potential or Berry connection, and $\varphi(k)$ is the Bloch wave function. We select the crossing points on the W-$\Gamma$ and L-U paths, and integrate the Berry vector potential around them over a anticlockwise closed $k$ path. We find that the Berry phase for two types of nodal loops are all $\pi$, which demonstrates that they are topologically nontrivial. Intriguingly, all nodal loops in reciprocal space are not isolated, but linked together by the common crossing points that locate on the path of $\Gamma$-X and $\Gamma$-K path to form a nodal chain structure, as illustrated in Fig. 4(c) and 4(d). Therefore, these two materials are spin-polarized nodal-chain semimetals.\\
\indent Topologically protected nodal loops generally result in the appearance of nontrivial surface states. We plot the projection of the nodal networks onto the (001) surface, as shown in Fig. 5(a). Figure 5(b) shows the equienergy slice at the Fermi level for the (001) surface states of Li$_2$MnO$_4$. We can observe several drumhead-like surface states originating from the four connecting points [the purple dots in the Fig. 5(b)] in the $\Gamma$-K. Besides, we also choose a path that pass through two crossing points in the $\Gamma$-X [cut in Fig. 5(a)] to calculate the surface states, as shown in Fig. 5(c). The two projected crossing points and the drumhead-like surface states are clearly visible. The similar drumhead-like surface states for Na$_2$MnO$_4$ are shown in Fig. S2.\\
\section*{\uppercase\expandafter{\romannumeral6}. THE EFFECTS OF STRONG CORRELATION, STRAINS AND SOC}
\indent Next, to clarify the robustness of the fully spin-polarized nodal chains, we calculated the electronic band structure of Li$_2$MnO$_4$ under different electronic correlations of d orbitals. Figure 6(a) shows the evolution of VBM and CBM (spin-down) at different U values ranging from 0 to 6 eV. We can see that the gaps always exist and grow as the U values increase. For the spin-up channel, we find that the crossings around the Fermi level have been preserved [see the bands in Fig. 6(b)]. To investigate the effect of lattice strains, we show how the gaps of the spin-down bands evolve under different biaxial strains, as shown in Fig. 6(c). The gaps are still well maintained. Figure 6(d) only shows the spin-up bands at 5$\%$ tensile and -5$\%$ compressive strains. We can see the crossing points still exist. Likewise, the related calculations of the Na$_2$MnO$_4$ compound are shown in Fig. S3, and the results are consistent with those of Li$_2$MnO$_4$. Finally, we'll discuss how SOC affects their electronic states. Spin is not a good quantum number in the presence of SOC, and the bands cannot be distinguished simply by spin, as shown in Figs. S4(b) and S4(d). The gaps are induced by SOC for all crossing points. However, they are less than 7.9 meV, and can be ignored at room temperature.\\
\section*{\uppercase\expandafter{\romannumeral7}. CONCLUSION}
\indent In summary, based on the first-principles calculations, we predict spinel X$_2$MnO$_4$ (X= Li, Na) have a Weyl-type nodal chain near the Fermi level. These two materials are dynamically, mechanically stable and have high cohesive energies. Further symmetry analysis reveals that the nodal chains are protected by the glide operation $\tilde{\mathcal{M}}_{z}$, mirror operation ${\mathcal{M}}_{\overline101}$ and their equivalent operations. Their topological features are proved by the nontrivial drumhead surface states. In addition, we found that the nodal chains are robust to electronic correlations and external lattice strains, and negligible band gaps are opened for the crossing points around the Fermi level with SOC. The study of these two materials not only broaden the scope of research into the interplay of topology and magnetism, but also provide more material choices for the future applications of spintronics.\\
\begin{acknowledgments}
The authors Ruirong Kang and Shenda He contributed equally to this work. This work is supported by the National Natural Science Foundation of China (Grant Nos. 11804287 and 11574260), Hunan Provincial Natural Science Foundation of China (2019JJ50577, 2021JJ30686, 2019JJ60006).
\end{acknowledgments}
\bibliography{Ref} 
\bibliographystyle{rsc} 
\end{document}